\documentclass[a4paper,11pt]{article}
\pdfoutput=1 

\usepackage{jinstpub} 
\usepackage{rotating}

\title{\boldmath A Novel Silicon Photomultiplier Readout Architecture for Low-light Applications}

\author[a,1]{V.~Cianciolo\note{Corresponding author.}}
\author[a,b]{Yu.\,V.~Efremenko,}
\author[a]{L.~Fabris,}
\author[b]{S.\,K.~Imam,}
\author[a]{S.\,I.~Penttila,}
\author[a]{J.\,C.~Ramsey,}
\author[c,2]{and R.~Santos Estrada\note{Currently at University of Connecticut \newline \hspace*{4.5mm}$^{\dag}$This manuscript has been authored by UT-Battelle, LLC, 
under contract DE-AC05-00OR22725 with the US  Department of Energy (DOE). The US government retains and the publisher, by accepting the article
for publication, acknowledges that the US government retains a nonexclusive, paid-up, irrevocable, worldwide license to publish or reproduce the published 
form of this manuscript, or allow others to do so, for US government purposes. DOE will provide public access to these results of federally sponsored research 
in accordance with the DOE Public Access Plan \href{http://energy.gov/downloads/doe-public-access-plan}{(http://energy.gov/downloads/doe-public-access-plan).}}}
\affiliation[a]{Oak Ridge National Laboratory, Oak Ridge, Tennessee, USA}
\affiliation[b]{University of Tennessee, Knoxville, Tennessee, USA}
\affiliation[c]{Berea College, Berea, Kentucky, USA}

\emailAdd{cianciolotv@ornl.gov}

\abstract{In this article we describe the photon detection readout electronics for the nEDM@SNS experiment. The chosen "photon counting" architecture, 
which utilizes high-efficiency silicon photomultipliers (SiPMs) and is appropriate for low-light applications, allows the use of a relatively high SiPM operating 
voltage. This maximizes photon detection efficiency and minimizes gain/efficiency voltage-dependence while eliminating direct optical cross-talk.}


\arxivnumber{} 

\begin{document}
\maketitle
\flushbottom

\section{Introduction}\label{sec:intro}


The nEDM@SNS experiment~\cite{edmjinst}, organized around principles laid out by Golub and Lamoreaux~\cite{gollam}, aims to measure the 
neutron electric dipole moment with unprecedented precision \linebreak ($\sigma_d < 3\times10^{-28}\,{\rm e \cdot cm}$) in order to shed light on the 
source of charge-parity violation responsible for the generation of matter in the Universe immediately following the Big Bang~\cite{sakharov, psw}. 

In this experiment ultracold neutrons are created and stored in a material trap (a "measurement cell") and exposed to highly uniform magnetic and electric fields. 
A non-zero electric dipole moment would alter the neutron precession by an amount proportional to the electric field. 
Neutron spin analysis is achieved by measuring the highly spin-dependent~\cite{n3He1,n3He2,n3He3} rate 
of \linebreak $n+\,^3{\rm He} \rightarrow p+t+764\,\rm{keV}$ capture events. The resulting proton and triton produce 
$\approx$\,5,000 extreme ultraviolet (EUV; 80\,nm) scintillation photons~\cite{euv1, euv2, lightvshv1, lightvshv2, lightvshv3} when their energy is deposited in 
the surrounding superfluid helium bath. The EUV photons are converted to optical photons with a deuterated 
tetraphenyl butadiene coating on the surface of the measurement cell walls~\cite{tpb1,tpb2}, which are subsequently captured in 
wavelength-shifting fibers, and guided to an array of silicon multipliers (SiPMs) and associated readout electronics housed in a temperature-controlled 
chamber at $\approx$-80\,C outside the experiment's magnetically shielded enclosure. 
Given sufficient stability in the light collection efficiency the number of detected photons can provide an {\it in situ} measurement of the volume-averaged electric field.
Details of the nEDM@SNS light collection system and simulations of its expected performance can be found in~\cite{devon}. 

Independent of SiPM choice, the SiPM readout architecture described here, appropriate for low-light applications, maximizes photon detection efficiency ($\epsilon_{\gamma}$), minimizes 
voltage and temperature dependence of $\epsilon_{\gamma}$ and gain, and eliminating direct optical cross-talk. 


\section{Design Considerations}\label{sec:considerations}

{\it Energy Resolution:} The $n+^3$He capture signal is monoenergetic and thus produces a fairly sharp peak in the number of photons detected per event. In contrast, background events%
\footnote{The primary background source is neutron $\beta$-decay. Other sources include 
Compton-scattered $\gamma$-rays (originating from neutron activation of nearby materials as well as ambient sources) and cosmic rays.} 
have fairly broad distributions. To achieve the target sensitivity, capture events must be detected with high 
efficiency ($\epsilon_{n^3{\rm He}}>0.93$) and the majority of $\beta$-decay background events must be rejected ($\epsilon_{\beta} < 0.5$)~\cite{edmjinst, devon}. 
Rejection of background events is achieved by applying a window cut on the signal peak in the detected photon spectrum, the 
effectiveness of which is set by the energy resolution. 


{\it Gain Stability: }Liquid helium scintillation photon yield is reduced in the presence of an electric field which can prevent electrons 
and ions (produced by ionizing energy loss) from radiative recombination~\cite{lightvshv1,lightvshv2,lightvshv3}. As a consequence, 
a precision measurement of the photon yield provides an elegant {\it in situ} measurement of the average electric 
field~\cite{gainmonitor}. At the design operating field ($E = 75$\,kV/cm) the scintillation yield dependence on the field can be approximated
by $\Delta PE/PE = -0.18 \Delta E/E$. Thus, to measure the electric field with 1\% accuracy requires measuring the scintillation yield with 0.18\% accuracy. 
Long-term gain drifts (for instance, due to degradation of the WLS measurement cell coating) can be separated from electric field drifts through periodic field-off gain calibration runs.

{\it Dark Rate: }Thermal generation of charge carriers in the photosensors (dark rate) produces temporally random 
single-photoelectron-equivalent background events.
High dark rate can produce accidental coincidences. Accidental coincidences between multiple dark rate pulses increases the trigger rate and can produce false n+3He capture signals. Accidental coincidences between dark rate pulses and scintillation events within the characteristic time of prompt EUV photon emission degrades the energy resolution. Accidental coincidences between dark rate pulses and scintillation events within the characteristic time of non-prompt EUV emission reduces background rejection capability~\cite{afterpulses}, and this sets the most stringent requirement of the rate ($D < 4$\,kHz/cell).

\begin{sidewaysfigure}[htbp]
\centering 
\includegraphics[width=\textwidth]{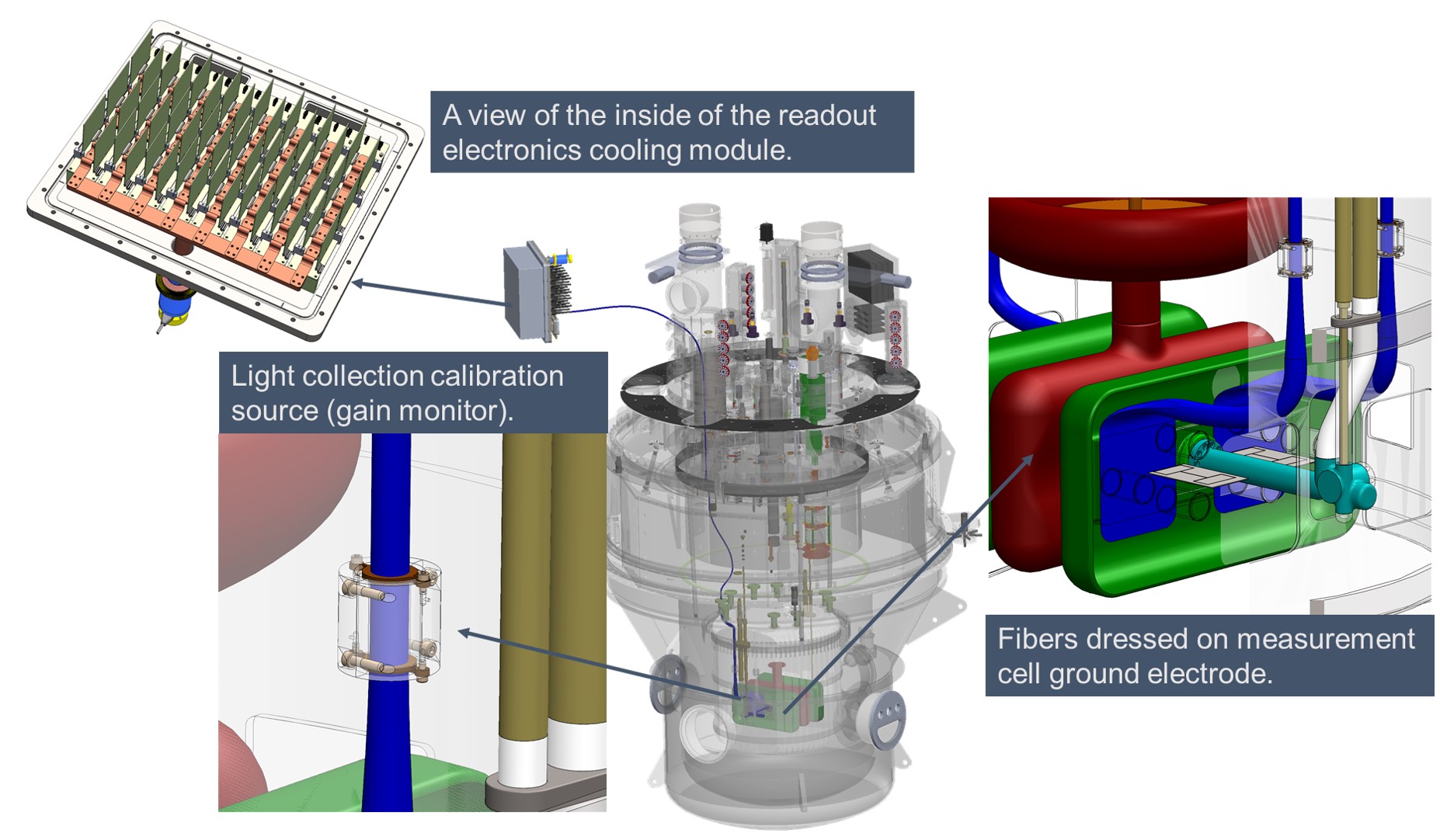}
\caption{\label{fig:geometry} Geometry of the nEDM@SNS light collection system. EUV photons are produced in superfluid helium in the measurement cells 
at $T \approx\,0.4$\,K and $E>40$\,kV/cm. A deuterated walvelength-shifting (WLS) coating on the measurement cell walls 
converts the EUV photons to blue so that they can be captured by WLS-doped plastic optical fibers. 
Optical gain is monitored by pulsing the fibers near the cell with a calibrated blue light source 
(an LED, with light delivered from room temperature via optical fiber, and/or an {\it in situ} source utilizing an $\alpha$ emitter and scintillator paint). 
Green photons produced in the WLS fibers are transported several meters 
to the photosensors located in a temperature-controlled enclosure outside the Faraday cage and Magnetic Shield Enclosure.}
\end{sidewaysfigure}

{\it Separation of Photon Generation/Detection: } The EUV scintillation photons are produced in a challenging environment ($T \approx 0.4$\,K, $E>40$\,kV/cm). 
In addition, the region near the measurement cell has strict non-magnetic requirements and every penetration of the outer vacuum vessel 
(which serves as a Faraday cage), especially by a condutor, represents a potential noise source that can interfere with the experiment's SQUID magnetometers~\cite{SQUID}. 
The closest acceptable location for the photosensors is several meters distant from measurement cells, outside the Faraday cage and the 
Magnetic Shield Enclosure, see Figure~\ref{fig:geometry}.

\section{Design and Implementation Details}

{\it Overview: } The long distance and tortuous path between photon generation and detection led to the choice of optical fiber light guides. 
After being converted to visible wavelengths the scintillation light is collected inside 192\,0.5 mm diameter wavelength-shifting (WLS) fibers pressed against the measurement cell wall adjacent to the ground electrode.
Because of the relatively short transmission length of WLS fibers (compared to clear, undoped fibers) both ends of each WLS fiber are mated to clear fibers close to the measurement cell.

The light guides transport the photons to an array of Silicon Photomultipliers (SiPMs) - one SiPM per WLS fiber end. 
Energy resolution places a premium on photon detection efficiency ($\epsilon_{\gamma}$), making SiPMs a natural photosensor 
choice~\cite{sipmoverview}. This is especially true given that a fiber is smaller than an individual SiPM, avoiding any inter-SiPM geometric inefficiency. 

We follow Otte {\it et al.}~\cite{nepomuk} and characterize SiPM performance as a function of relative over-voltage 
($R_{ov} = (V_{op} - V_{bd}) / V_{bd}$), where $V_{op}$ is the operating voltage and $V_{bd}$ is the breakdown voltage. 
This variable is convenient because the probability of breakdown can be described using a single parameter: $P_{bd}(R_{ov}) = 1-\exp(-R_{ov}/\alpha)$.

Photon detection efficiency ($\epsilon_{\gamma}$) increases with $R_{ov}$. $\epsilon_{\gamma}$ dependence on $R_{ov}$ (and thus temperature) decreases with $R_{ov}$  
(the two are closely related by the strong temperature dependence of $V_{bd}$, and 
thus $R_{ov}$ when $V_{op}$ is fixed).
These benefits must be weighed against two negative effects which both increase with $R_{ov}$: dark rate (for which the most stringent limit is set by a desire to measure 
non-prompt scintillation photons which may provide improved rejection of background events) 
and direct optical cross-talk (which works against the energy resolution improvement).

Dark rate can be easily reduced by cooling the SiPMs. Direct optical cross-talk is eliminated with a "photon counting" readout architecture which uses a 
non-retriggerable voltage discriminator to produce a uniform voltage pulse when a SiPM is struck. Single photoelectrons can be cleanly identified above the electronic noise pedestal, so
single photoelectron signals can be selected with $\approx100\%$ efficiency. The discriminator is dead for the duration of its output pulse (chosen to be $\approx 50$\,ns), 
so direct optical cross-talk ignored since direct optical photons occur immediately following the primary photon interaction in the SiPM. Delayed optical cross talk and SiPM afterpulsing 
events can occur up to 100\,ns after the primary photon interaction in the SiPM, but most of these events are also eliminated by the dead window. 
Those which remain will not contribute to the determination of the energy of the prompt scintillation pulse.
Those that remain can be clearly separated from the non-prompt scintillation light due to their short time constant.
The dead window introduces inefficiency when multiple scintillation photons from the same
 event hit the same SiPM, but in the prevailing low-light conditions this rarely happens.
The photon counting readout architecture also has the benefit of eliminating any impact from SiPM gain variation.%
\footnote{SiPM gain variation can result from sensor-to-sensor differences in intrinsic gain, $V_{bd}$ or temperature. 
By producing a digital output  "photon counting"  ignores these variations, assuming the gain is high enough that a 
single photoelectron is well above the discriminator threshold.}
Signals from 16 SiPMs are multiplexed into a single readout channel whose output is a voltage pulse with amplitude proportional to the number of photons detected on those SiPMs. 
Thus each channel provides a count (from 0 to 16) of the number of struck SiPMs. Note, for an average signal we expect $\approx 1$ struck SiPM per channel.

{\it Fiber choice: }
Kuraray Y-11 fibers~\cite{kuraray} have been thoroughly characterized by multiple experimental groups~\cite{nova, mu2e, majorana}. 
The Y-11 absorption spectrum  has good overlap with the TPB emission spectrum~\cite{tpb1,tpb2}, and the Y-11 emission spectrum is 
peaked near 500\,nm, well-suited for many SiPMs. Double-cladded fibers have superior capture fraction (10.8\%). The S-type is resistant 
to cracks (minimum bend diameter $\approx 100 \times$~fiber diameter) and we have observed no degradation in performance following 
numerous thermal cycles. Simulations~\cite{devon} show standard dye density (200\,ppm) and fiber diameter (either 0.5 or 1\,mm) are appropriate.

We transition to clear fibers (ESKA SK-40, 1\,mm diameter~\cite{eska}) to maximize transmission over the relatively long distance ($\approx 9$\,m) to the photosensors.

{\it Photosensor choice: } Various types of silicon sensors have high photon detection efficiency. SiPMs were chosen due to their conveniently low 
operating voltage ($\approx 50$\,V) and superior energy resolution which allows clean identification of single photoelectrons. Otte {\it et al.}~\cite{nepomuk} 
carried out extensive performance characterizations of SiPMs from FBK (NUV-HD), Hamamatsu (S13360-3050CS), and SensL (J-Series 30035). 
Hamamatsu sensors were shown to have higher  $\epsilon_{\gamma}^{max}$ (especially at green wavelengths), a wider $\epsilon_{\gamma}$ vs.~$V_{op}$ plateau, and lower dark rate. 
We selected the Hamamatsu S13360-1375CS~\cite{hamamatsu} - a different package of the same family of sensors tested in~\cite{nepomuk} with smaller 
area and larger pixel size. The smaller area (1.3\,mm $\times$ 1.3\,mm) is optimal for reading out a single fiber, and the larger pixel size (75\,$\mu$m) 
increases $\epsilon_{\gamma}$ 20\% due to increased geometric efficiency (the resulting reduction in dynamic range is irrelevant at our low light levels). 

One concern with the Hamamatsu sensors was the large range of $V_{bd}$ ($\pm 5$\,V) listed in the product datasheet. 
Without a separably tunable $V_{op}$ for each SiPM, $V_{bd}$ variations result in $R_{ov}$ variations that lead to 
 reduced $\epsilon_{\gamma}$ (in sensors with higher $V_{bd}$) and higher dark rate (in sensors with lower $V_{bd}$).
Sensor-dependent $\epsilon_{\gamma}$ values also worsen the signal peak resolution.

Otte {\it et al.}~\cite{nepomuk} showed that $\epsilon_{\gamma}$ as a function of $R_{ov}$ is well described by
\begin{equation}\label{eq:PDE}
\epsilon_{\gamma}/\epsilon_{\gamma}^{max}(R_{ov}) = (1-\exp(-R_{ov}/\alpha))
\end{equation}
where $\alpha \approx 0.05$ for Hamamatsu sensors exposed to photons between 500-589\,nm.
The expected mean number of photoelectrons in a $n+^3$\,He signal event is $\approx$17~\cite{devon}, giving a shot-noise limited signal peak resolution of 
$\sigma_{min}/\mu \approx 1/\sqrt{17} = 0.24$. To prevent sensor variations in $V_{bd}$ from contributing significantly to the signal peak resolution 
we require sensor-to-sensor $\epsilon_{\gamma}$ variations to be an order of magnitude smaller than this: $d(\epsilon_{\gamma}/\epsilon_{\gamma}^{max})/dR_{ov} \times \Delta R_{ov} = \frac{(1-\epsilon_{\gamma}/\epsilon_{\gamma}^{max})}{\alpha} \times \Delta R_{ov} \le 0.024$. 
At an operating voltage such that $\epsilon_{\gamma}/\epsilon_{\gamma}^{max} = 95\%$,
$R_{ov} = 0.15$ and  $d(\epsilon_{\gamma}/\epsilon_{\gamma}^{max})/dR_{ov} = 1$. For the 
Hamamatsu sensors $V_{bd} \approx 50$\,V, leading to the requirement that  $\Delta V_{bd} \le 1.2$\,V%
\footnote{Otte {\it et al.}~\cite{nepomuk} found $dV_{bd}/dT \approx 55$\,mV/C for the Hamamatsu sensors, so the $V_{bd}$ 
uniformity requirement leads to a modest temperature uniformity requirement: $\Delta T < 22.7$\,C.}
and making $\pm 5$\,V variations unacceptable. 
However, Hamamatsu guaranteed and delivered 1,000 sensors within $\pm 0.75$\,V of the nominal $V_{bd}$ at no extra cost.%
\footnote{The quoted value is presumably a batch-to-batch variation rather than variation within a batch.} 
It follows immediately that such variations do not contribute significantly to efficiency loss. Otte {\it et al.}~\cite{nepomuk} studied the dark rate dependence  on both
$R_{ov}$ and temperature. Sensors with $V_{bd}$ 0.75\,V below nominal will have $R_{ov} \approx 0.015$ above nominal, corresponding to a 
15\% increase in the dark rate. This modest increase is easily mitigated by reducing the SiPM temperature $\approx 1.3$\,C.

{\it Operating temperature: }
Otte {\it et al.}~\cite{nepomuk} found that the Hamamatsu sensor dark rate was $D \approx 0.25$\,kHz/mm$^2$ at $R_{ov} = 0.15$, 
and T= -40\,C and had a strong temperature dependence: at fixed $R_{ov}$, $D(T) = D(T_0)\times 2^{\frac{T-T_0}{7.0{\rm C}}}$.%
\footnote{Dark rate temperature dependence in Figure~23 of Otte {\it et al.} is plotted vs.~$V_{op}$. 
Dependence vs. $R_{ov}$ is obtained from Equation 13 and fit parameters given in Table 2 of the same reference.}
We have 384 sensors per cell, for a total sensor area of $\approx 650$\,mm$^2$, so for $D < 4$\,kHz/cell we require $T<-77$\,C. 
With this dark rate the contribution of false hits to the prompt scintillation pulse (width < 100\,ns) will be 
negligible ($4\times 10^{-4}$), as will the false trigger rate (1.6\,Hz, assuming a simple double coincidence).

{\it Voltage and temperature stability: } 
The photon counting technique reduces the system gain ($G$ = charge per photon) stability requirement to a photon detection efficiency stability requirement
since each detected photon outputs the same charge: $dG/G = d\epsilon_{\gamma}/\epsilon_{\gamma} \le 0.18\%$. 
In the discussion following Equation~\ref{eq:PDE} we showed that at our operating voltage 
$d(\epsilon_{\gamma}/\epsilon_{\gamma}^{max})/dR_{ov} = 1$, from which the voltage stability requirement can be derived: 
$\Delta R_{ov} \le 0.18\%$, or $\Delta(V_{op} - V_{bd}) \le 90$\,mV, since $V_{bd} \approx 50$\,V. 
This imposes requirements on both  $V_{op}$ and $ V_{bd}$. Commercially available supplies~\cite{BKPrecision} easily achieve the required $V_{op}$ stability. 
The $V_{bd}$ stability requirement is effectively a temperature stability requirement: $\Delta T \le 1.6$\,C.

It is worth noting that without photon counting the system gain stability requirement applies to the product of the photon detection efficiency and the SiPM analog gain ($G_A$). 
Otte {\it et al.}~\cite{nepomuk} showed $G_A \propto (V_{op}-V_{bd})$ for the Hamamatsu sensors.
So, to achieve $dG_A/G_A \le 0.18\%$ would require $d(V_{op}-V_{bd})/(V_{op}-V_{bd}) \le 0.18\%$. 
With  $V_{bd} ~ 50$\,V and $R_{ov} ~ 0.15$ this gives $\Delta(V_{op} - V_{bd}) \le 13.5$\,mV, 
significantly more stringent than required with the photon counting technique.

\begin{figure}[htbp]
\centering 
\includegraphics[width=4in]{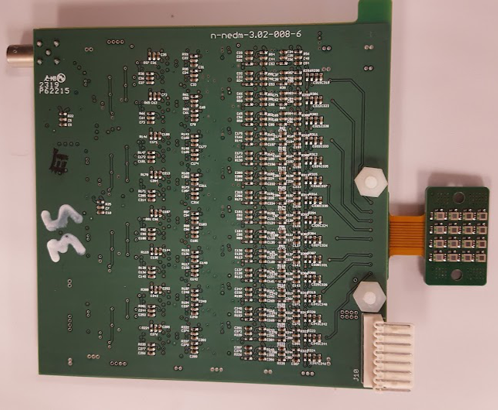}
\caption{\label{fig:boardphoto}Photo of a 16-channel SiPM host board connected via rigid-flex circuitry to its signal processing board.}
\end{figure}

{\it Circuit details: }
16 SiPMs are mounted on a rigid-flex circuit and connected to a $4" \times 4"$ PCB with the signal processing circuitry (see Figure~\ref{fig:boardphoto}). 
The flex circuit connection between the two boards simplifies the cooling geometry, as described below.

\begin{figure}[htbp]
\centering 
\includegraphics[width=\textwidth]{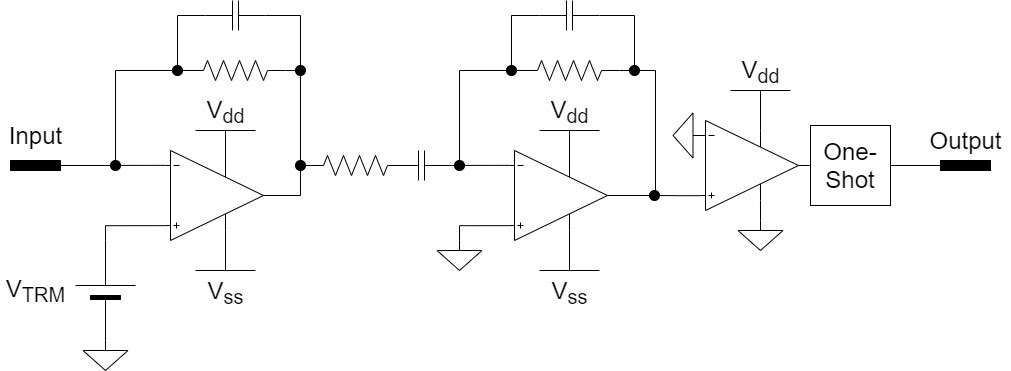}
\caption{\label{fig:circuit}Simplified schematic of one readout channel. Two stages of amplification (the second is AC coupled to remove the DC value of the baseline), a comparator, and a one-shot are all that is needed to reliably read out our SiPMs. }
\end{figure}

Each SiPM has a dedicated readout circuit (see Figure~\ref{fig:circuit}), the function of which is to identify whether a photon was detected by the SiPM. This circuit
takes advantages of the extremely high signal-to-noise ratio resulting from the small area and low-capacitance of the SiPMs, which allows use of a commercial operational amplifier as 
a front-end device without excessive concern about electronic noise. The operational amplifiers were chosen based on low noise, low power, adequate bandwidth, 
and the ability to correctly function at cryogenic temperatures. 
Texas Instruments’ OPA~836~\cite{TI-OPA836} has a quiescent current of only 1\,mA from a symmetric 2.5\,V rail and yet has a noise of 4.6\,${\rm nV}/\sqrt{\rm Hz}$. 
The dynamic response is outstanding with a 205\,MHz bandwidth and a fast rise time with rail-to-rail outputs. 
This component is specified for an operating range of -40\,C to +125\,C, but successful tests were performed at liquid nitrogen temperature (77\,K) establishing good performance at cryogenic temperatures. 
The use of an operational amplifier as the front-end element has the added advantage of allowing fine trimming of the SiPM bias (when DC coupled, as is our case) by  
shifting the DC value at the non-inverting port.\! 
\footnote{See voltage source V$_{\tt TRM}$ in Figure~\ref{fig:circuit}. Since the sensors delivered by Hamamatsu had sufficiently uniform $V_{bd}$ we did not need to take advantage of this feature.}
Photon detection is registered by a low-power comparator with a fixed threshold and proper hysteresis. Once the signal presence 
is detected a custom, fast, non-retriggerable one-shot shapes its duration to a selectable width. The reason for a custom one-shot is that commercial devices do not reliably allow output pulse widths less than 75\,ns. 
In order to minimize dead time associated with the photon counting process we required a pulse width of 50\,ns. The digital TTL response of the 16 one-shots on a processing board are converted to a pre-determined 
analog amplitude, and summed in an analog fashion into an inverting amplifier that provides a single output for all 16 SiPMs. It should be noted that, if signal rates are low enough this architecture allows each SiPM, or 
group of SiPMs, to be effectively encoded in both pulse duration and pulse amplitude to provide unequivocal identification of struck SiPMs if needed. 

\pagebreak

\begin{figure}[htbp]
\centering 
\includegraphics[width=\textwidth]{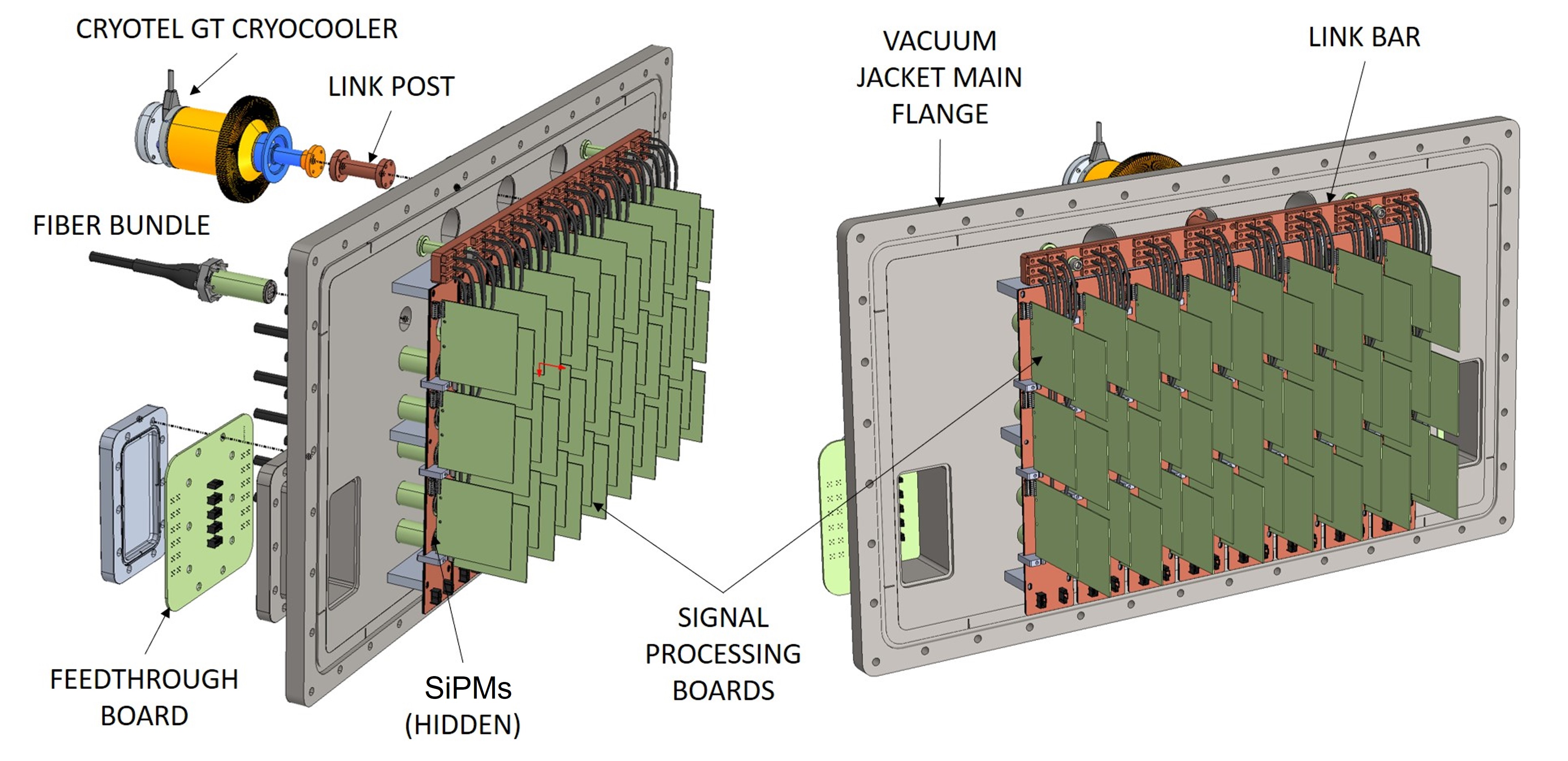}
\includegraphics[width=\textwidth]{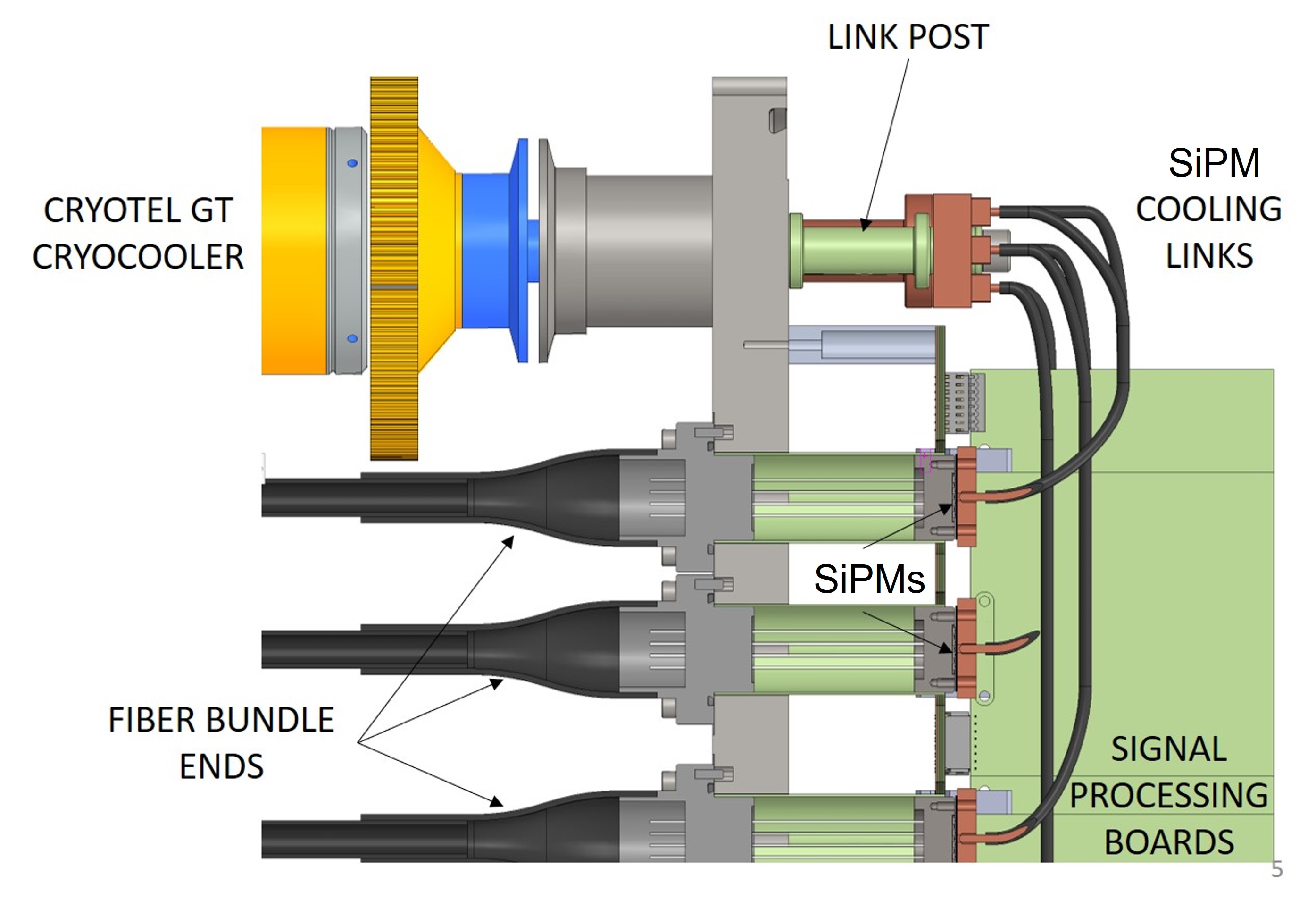}
\caption{\label{fig:cooling}Geometry and nomenclature of the readout electronics cooling system.}
\end{figure}

\begin{figure}[htbp]
\centering 
\includegraphics[width=\textwidth]{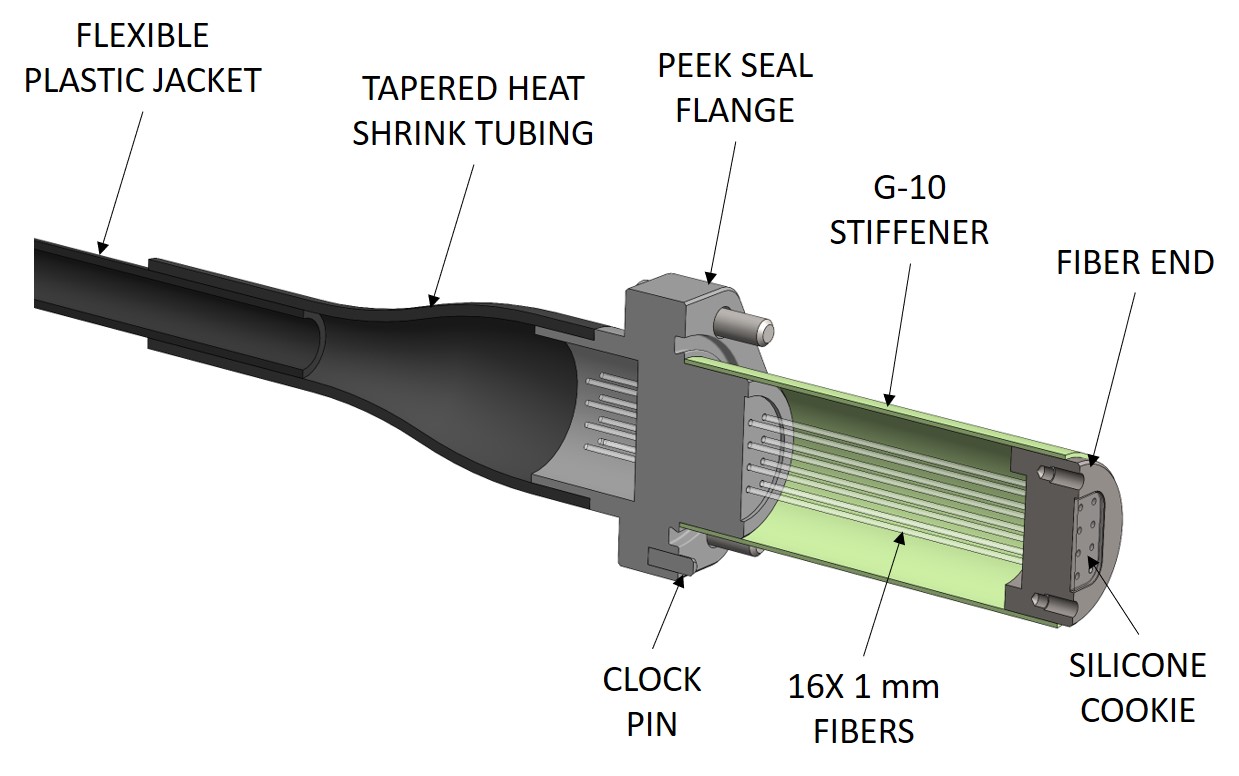}
\caption{\label{fig:fiberbundle} Design of the low-heat conductivity fiber bundle interface to the SiPMs. }
\end{figure}

{\it Cooling details: }
The goal of the SiPM readout cooling system is to keep the SiPMs at or below -80\,C, with temperature stability on the order of 1-2\,C and uniformity within $\pm 10$\,C. 
To achieve these goals the SiPMs are mounted in a vacuum enclosure and thermally anchored to a 16\,W Cryotel GT cryocooler~\cite{Cryotel}.
The basic geometry of the cooling system is shown in Figure~\ref{fig:cooling}. 
The primary conductive heatloads to the SiPMs are from the associated fiber light guide (0.11\,W) and processing board (0.14\,W). 
The heatload from the fiber light guide is minimized by the design shown in Figure~\ref{fig:fiberbundle} which limits the area of the conductive pathway to that of the fibers themselves plus that of the G-10 stiffener.
In addition to the $48 \times 0.25=12$\,W conductive load the cryocooler must remove an additional 5.8\,W, primarily due to radiative heating.
The thermal performance of the system has been modeled with ANSYS, with pressed joint conductances calculated using a rough model from~\cite{Barron}
Calculations show a comfortable amount of extra cooling capacity and temperature uniformity well within the $\pm 10$\,C goal. 
The temperature uniformity requirement translates to a heatload uniformity requirement.
Increased heatload results in increased temperature of the cryocooler heat exchanger and reduced temperature drops across the thermal links.
Given the cryocooler cooling curve and the design thermal link conductances one finds $\Delta T/\Delta Q = 6.5$\,$^{\circ}K\!/{\rm W}$, so the heatload must be stable to within $\pm0.2$\,W.
The heatloads, and therefore the temperature, are expected to be quite constant (the vacuum enclosure is in a temperature controlled experimental hall and the signal processing board current draw is fixed).
But the system will be instrumented with two 25\,W resistive heaters and thermometry on a subset of SiPM boards, so active control can be implemented if needed.

\pagebreak

\section{Performance}

The performance of the SiPMs and associated readout electronics was evaluated by mounting each 16-channel readout unit in a cold box (Delta 9023) maintained at -60\,C. 
The SiPMs were illuminated by Kuraray Y-11 fibers, each of which could be independently pulsed with an LED. 
The pulse size was set so that the average number of photons per pulse was much less than one to guarantee that all effectively every signal above the pedestal corresponded to a single photoelectron.
Both the digital output and a diagnostic analog output from each readout unit were fed into a CAEN DT5730B waveform digitizer.

\begin{figure}[htbp]
\centering 
\includegraphics[width=\textwidth]{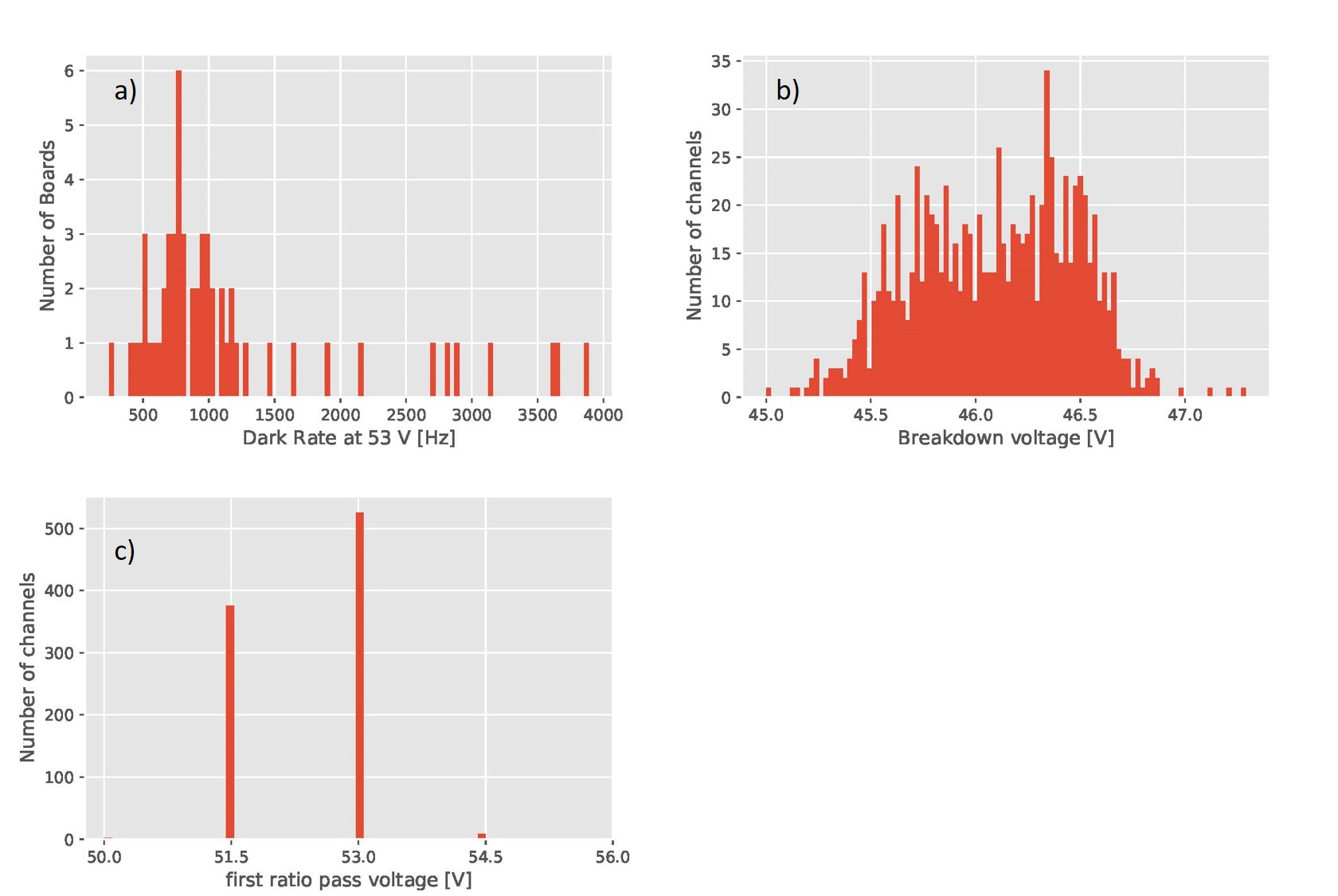}
\caption{\label{fig:results} {Panel a) shows the dark rate of each 16-channel readout unit at $V_{bd} \approx 0.15$ and $T\approx -60$\,C. Panel b) shows the estimated breakdown voltage of each individual SiPM. Panel c) shows the voltage where the digital output has efficiency $> 95$\%. }}
\end{figure}

The dark rate was measured for each readout unit, as shown in Figure~\ref{fig:results}a. 
The average rate per board (total SiPM area $= 16\times1.3\,{\rm mm}\times 1.3\,{\rm mm} = 27\,{\rm mm}^2$) is $\approx 800$\,Hz, in line with expectations from Otte {\it et al.}~\cite{nepomuk} at this temperature. 
Outliers may be due to imperfections in screening ambient light. If not, there are sufficient spares to avoid using the noisier SiPMs.
By running at somewhat lower temperature (-77\,C)  we can meet the desired dark rate of 4\,kHz/cell (twenty-four 16-channel readout units).

The breakdown voltage for each SiPM, plotted in  Figure~\ref{fig:results}b was determined by pulsing individual SiPMs at a variety of $V_{op}$ settings,
 integrating the resulting analog pulses, plotting the average single photoelectron signal size ({\it i.e.}, the gain) vs. $V_{op}$, and finding the $x$-intercept of a linear fit to the results. 
All but a few sensors are within Hamamatsu's $\pm 0.75$\,V guarantee. All are well within our $\pm 1.2$\,V requirement.

Finally the performance of the discriminator circuit was evaluated by testing for the presence of a digital signal when a single photoelectron analog signal was present. 
Figure~\ref{fig:results}c shows the distribution of $V_{op}$ settings where the boards first achieved 95\% efficiency on the digital signal. 
All but one board had full efficiency at or below the nominal  $V_{\tt bias}$ setting (53\,V).

\section{Conclusions}

We have developed and successfully tested nearly 1,000 channels of Silicon Photomultiplier readout electronics for the nEDM@SNS experiment.
The "photon counting" architecture provides many advantages for low-light applications and the design is generic
enough that it may find use in other similar applications.

\acknowledgments

We gratefully acknowledge the support of the U.S. Department of Energy Office of Nuclear Physics through grant DE-AC05-00OR22725.



\end{document}